\documentclass[a4paper,12pt]{article}
\usepackage{amsmath}
\usepackage{amssymb}
\usepackage{latexsym}
\usepackage{enumerate}
\newcommand{\be}{\begin{eqnarray}}
\newcommand{\ee}{\end{eqnarray}}

\newcommand{\bsub}{\begin{subequations}}
\newcommand{\esub}{\end{subequations}}

\newcommand{\disfrac}[1][2]{\displaystyle\frac}
\begin{document}

\title{Energy-Momentum Localization for a Space-Time Geometry Exterior to a
Black Hole in the Brane World}
\author{Irina Radinschi$^{\text{*1}}$, Theophanes Grammenos$^{\text{**2}}$ and Andromahi
Spanou$^{\text{***3}}$ \\
$^{\text{1}}$Department of Physics \\
``Gh. Asachi'' Technical University, \\
Iasi, 700050, Romania\\
$^{\text{2}}$Department of Civil Engineering, \\
University of Thessaly, 383 34 Volos, Greece\\
$^{\text{3}}$School of Applied Mathematics and Physical Sciences,\\
National Technical University of Athens, 157 80, Athens, Greece\\
$^{\text{*}}$radinschi@yahoo.com, $^{\text{**}}$thgramme@uth.gr,\\
$^{\text{***}}$aspanou@central.ntua.gr}
\date{}
\maketitle

\begin{abstract}
In general relativity one of the most fundamental issues consists in
defining a generally acceptable definition for the energy-momentum density.
As a consequence, many coordinate-dependent definitions have been presented,
whereby some of them utilize appropriate energy-momentum complexes. We
investigate the energy-momentum distribution for a metric exterior to a spherically symmetric black hole in the brane world by applying the Landau-Lifshitz and
Weinberg prescriptions. In both the aforesaid prescriptions, the energy thus
obtained depends on the radial coordinate,  the mass of the black hole and a parameter $\lambda _{0}$,
while all the momenta are found to be zero. It is shown that for a special
value of the parameter $\lambda _{0}$, the Schwarzschild space-time geometry
is recovered. Some particular and limiting cases are also discussed.
\end{abstract}

\section{Introduction}

In general relativity there has not  been given so far a generally accepted expression for the 
energy density in gravitational fields, while none of the various approaches used has provided a strong indication for its candidacy as the best for the energy-momentum localization. Over the past years, numerous attempts have been made in order to calculate the
energy-momentum distribution of the gravitational field, utilizing tools such as superenergy
tensors \cite{Bel}, quasi-local expressions \cite{Brown}, energy-momentum complexes \cite{Einstein}-\cite{Gadir}
and even the tele-parallel (tetrad) theory of gravity \cite{teleparallel}. 

A few remarks concerning the aforementioned different approaches are deemed necessary. The approach using the superenergy tensors has improved the energy-momentum expressions a great deal in recent years, while  the definitions of the quasi-local mass exhibit the advantage of being applicable to any coordinate system. On the other hand, pseudotensorial definitions which make use of the energy-momentum complexes
of Einstein \cite{Einstein}, Landau-Lifshitz \cite{Landau}, Papapetrou \cite{Papapetrou}, 
Bergmann-Thomson \cite{Bergmann}, M\o ller \cite{Moller_1}, Goldberg \cite{Goldberg},
Weinberg \cite{Weinberg} and Qadir-Sharif \cite{Gadir}  have been applied to many spacetime
geometries, yielding also significant results. Here it is worth mentioning that the Einstein, Landau-Lifshitz, Papapetrou and Weinberg (henceforth ELLPW) prescriptions use quasi-Cartesian coordinates, whereas the M\o ller energy-momentum complex can be applied
to any coordinate system. 
Finally, as far as the $3+1$, $2+1$ and $1+1$ dimensional space-times are concerned, we may point out how useful the pseudotensorial definitions have been proven  for the evaluation of the gravitational
energy-momentum. We may notice that for many
gravitational backgrounds different prescriptions have given the same expression
for the energy-momentum (e.g., \cite{Virbhadra} and references therein, mainly on the LL and W prescriptions). 

In the tele-parallel theory of gravity a regularized expression for
the gravitational energy-momentum is derived. It has been shown that the theory of general relativity can be reformulated in the context of the tele-parallel
(Weitzenb\"{o}ck) geometry. Recently, there has been an increasing interest in calculations employing this theory and many significant results for various space-times (\cite{teleparallel}) have been obtained. At this point, the similarity of some results generated by pseudotensorial prescriptions and their tele-parallel versions (see, e.g., \cite{Gamal_1}) should also be stressed.

However, even if the aforementioned approaches have not led to a generally accepted, well-defined expression for the gravitational energy density, they have contributed to 
establishing a basis for the performance of valid calculations.

The remainder of this paper is organized as follows: in Section 2
a short presentation of the energy-momentum complexes is given along with some details on the Landau-Lifshitz and
Weinberg energy-momentum complexes utilized for the calculations in the present paper. Section 3 is devoted to the calculations of the
energy-momentum for a new black hole solution in the brane world. Finally, the Discussion contains a summary of the
results and a presentation of some particular limiting cases. Throughout the paper we have used geometrized units ($c=1; G=1$) and the signature ($+,-,-,-$) for the Landau-Lifshitz and Weinberg prescriptions
in Schwarzschild-Cartesian coordinates, while Greek indices range from $0$ to $3$ while Latin
indices  from $1$ to $3$.

\section{Energy-Momentum Complexes - A Short Presentation}

The energy-momentum complexes of Einstein \cite{Einstein}, Landau-Lifshitz \cite{Landau},
Papapetrou \cite{Papapetrou}, Bergmann-Thomson \cite{Bergmann}, M\o ller \cite{Moller_1}, Goldberg \cite{Goldberg},
Weinberg \cite{Weinberg} and Qadir-Sharif \cite{Gadir} have been employed for numerous gravitational backgrounds, leading to acceptable results for the energy-momentum localization. An energy-momentum
complex is made up of three components describing  the gravitational
field, the matter and the possible non-gravitational fields, respectively. The
energy-momentum complexes conserve the differential conservation law.

Among the characteristic features, or better said, weaknesses of these complexes,
we should mention the fact that they are coordinate dependent. This,
together with the fact that the calculations can be performed only in
quasi-Cartesian coordinates, except for the M\o ller prescription, has led to their rather scarce utilization for energy-momentum localization.

Chang, Nester and Chen \cite{Chang} attempted to rehabilitate the energy-momentum
complexes, by showing that these are quasi-local and by emphasizing their
importance. One should point out that different quasi-local definitions correspond to different
boundary conditions. Also, So, Nester and Chen \cite{So}, in an attempt to improve the pseudotensorial
mechanism, showed that ``for small vacuum regions the Einstein, Landau-Lifshitz, Papapetrou, Weinberg, Bergmann-Thomson and Goldberg pseudotensors have the zero order material limit
required by the equivalence principle''. The M\o ller prescription though does not comply to this.

Furthermore, the aforementioned pseudotensors are not proportional to the
Bel-Robinson tensor. Thus, some attempts have been made in order to find a profitable
combination of different pseudotensors. This way, they managed to make
improvements and to elaborate an independent combination of Bergmann-Thomson, Papapetrou and Weinberg
energy-momentum complexes. They also formulated a one parameter set of
linear combinations of classical pseudotensors with the required Bel-Robinson
connection.

Part of the significance of the energy-momentum complexes relies on the proof  that several energy-momentum complexes ``coincide'' for any metric
of the Kerr-Schild class \cite{Aguirre}-\cite{Virb}. Furthermore, we may
notice that the definitions provided by Einstein, Landau-Lifshitz,
Papapetrou, Bergmann-Thomson, Weinberg and M\o ller agree with the quasi-local mass definition
introduced by Penrose \cite{Penrose} and developed by Tod \cite{Tod}, at least for some
space-times and some energy-momentum complexes. Moreover, it has been shown
that different prescriptions yield the same result for a given gravitational
background under the condition that the calculations are performed in Schwarzschild
Cartesian and Kerr-Schild coordinates, while satisfying 
results have also been obtained for $2$ and $3$ dimensional space-times
\cite{Virbhadra}, \cite{Aguirre}. However,  there do exist some cases where different energy-momentum complexes yield different results when the calculations are performed in
Schwarzschild Cartesian  and  Kerr-Schild coordinates \cite{Virb}.

Last but not least, we should mention two more viewpoints supporting the importance of the
energy-momentum complexes. Lessner \cite{Lessner} has argued
that ``the M\o ller definition is a powerful concept of
energy and momentum in general relativity'', while Cooperstock
\cite{Cooper} has emphasized, in his hypothesis of utmost importance, the fact that
``the energy and momentum are confined to the regions of
non-vanishing energy-momentum tensor for the matter and all
non-gravitational fields''.

Now, we shall present the energy-momentum complexes used in the present work.

The Landau-Lifshitz energy-momentum complex \cite{Landau} reads
\begin{equation}\label{LLemc}
L^{\mu \nu }=\frac{1}{16\pi }S_{,\,\rho \sigma }^{\mu \rho \nu \sigma } 
\end{equation}
and the corresponding superpotentials are given by: 
\begin{equation}\label{LLsp}
S^{\mu \nu \rho \sigma }=-g(g^{\mu \nu }g^{\rho \sigma }-g^{\mu \rho }g^{\nu
\sigma }). 
\end{equation}
$L^{00}$ and $L^{0i}$ represent the energy and the momentum density components,
respectively. The Landau-Lifshitz energy-momentum complex satisfies the
local conservation law
\begin{equation}\label{LLcl}
L_{,\,\nu }^{\mu \nu }=0. 
\end{equation}
The integration of $L^{\mu \nu }$ over the 3-space gives the expression for the
energy-momentum four-vector:
\begin{equation}\label{LL4p}
P^{\mu }=\int \int \int L^{\mu 0}\,dx^{1}dx^{2}dx^{3}.
\end{equation}
By using Gauss' theorem one obtains
\begin{equation}\label{LLgauss}
P^{\mu}=\frac{1}{16\pi}\int\int S^{\mu 0 i\nu}_{,\nu}n_{i}dS=\frac{1}{16\pi}\int\int U^{\mu 0 i} n_{i}dS.
\end{equation}

The Weinberg energy-momentum complex \cite{Weinberg} is given by
\begin{equation}\label{Wemc}
W^{\mu \nu }=\frac{1}{16\pi }D_{,\,\lambda }^{\lambda \mu \nu },
\end{equation}
where the superpotentials are 
\begin{equation}\label{Wsp}
D^{\lambda \mu \nu }=\frac{\partial h_{\kappa }^{\kappa }}{\partial
x_{\lambda }}\eta ^{\mu \nu }-\frac{\partial h_{\kappa }^{\kappa }}{\partial
x_{\mu }}\eta ^{\lambda \nu }-\frac{\partial h^{\kappa \lambda }}{\partial
x^{\kappa }}\eta ^{\mu \nu }+\frac{\partial h^{\kappa \mu }}{\partial
x^{\kappa }}\eta ^{\lambda \nu }+\frac{\partial h^{\lambda \nu }}{\partial
x_{\mu }}-\frac{\partial h^{\mu \nu }}{\partial x_{\lambda }},
\end{equation}
with 
\[
h_{\mu \nu }=g_{\mu \nu }-\eta _{\mu \nu } 
\]
and $W^{00}$, $W^{0i}$ represent the energy and the momentum density
components, respectively. The Weinberg energy-momentum complex satisfies the
local conservation law
\begin{equation}\label{Wcl}
W_{,\,\nu }^{\mu \nu }=0.
\end{equation}
The integration of $W^{\mu \nu }$ over the 3-space yields the expression for the
energy-momentum four-vector:
\begin{equation}\label{W4p}
P^{\mu }=\int \int \int W^{\mu 0}\,dx^{1}dx^{2}dx^{3}.
\end{equation}
By applying Gauss' theorem we obtain
\begin{equation}\label{Wgauss}
P^{\mu }=\frac{1}{16\pi }\int \int D^{i0\mu}n_{i}dS.
\end{equation}

\section{Energy-Momentum for a  New Black Hole Solution in the Brane World}

Since the introduction of the brane world notion and the role of gravity in it (for a review see, e.g., \cite{Maartens} and references therein), an increasing interest in black hole solutions in this context has led to an accordingly increasing number of relevant works, such as \cite{Tanahasi} just to select but a few publications that review the subject. On the other hand, there is rather little work concerning the calculation of energy-momentum in brane world models, see, e.g., \cite{Liu} and in particular for black hole solutions in such models \cite{Gamal_2}.

Searching for black hole solutions on the brane, Casadio, Fabbri and Mazzacurati \cite{Casadio} found a new static, spherically symmetric and asymptotically flat solution that has the line element:
\begin{equation}\label{casadio-line-element}
ds^{2}=(1-\frac{2M}{r})dt^{2}-\frac{(1-\frac{3M}{2r})}{(1-\frac{2M}{r})(1-%
\frac{\lambda _{0}}{r})}dr^{2}-r^{2}(d\theta ^{2}+\sin ^{2}\theta d\varphi
^{2}),
\end{equation}
where $\lambda_0 \in\mathbb{R}^{+}$ and $M$ is the ADM mass. In fact, it is shown \cite{Germani} that the line element (\ref{casadio-line-element}) can be also used to describe space-time geometry exterior to a homogenous star on the brane, while for $\lambda_0 > 2M$ it gives a wormhole  geometry \cite{Bronnikov}.
Furthermore, the metric described by (\ref{casadio-line-element}) represents 
 a black hole solution in the brane world also in the context of the Teleparallel Equivalent of General Relativity (TEGR) \cite{Gamal_2}. 
 
For the specific values $\lambda_0=0$, $M=0$ the space-time geometry given by (\ref{casadio-line-element}) is flat, while for $\lambda_0 =\frac{3}{2}M$, (\ref{casadio-line-element}) becomes the Schwarzschild line element.

The line element (\ref{casadio-line-element}) can be written in Schwarzschild-Cartesian coordinates, which we need for the utilization of the  Landau-Lifshitz and Weinberg prescriptions, as follows:
\begin{equation}\label{metricSC}
ds^{2}=B(r)dt^{2}-(dx^{2}+dy^{2}+dz^{2})-\frac{A(r)-1}{r^{2}}
(xdx+ydy+zdz)^{2}, 
\end{equation}
with $A(r)$ and $B(r)$ given by
\begin{equation}\label{A+B}
A(r)=\frac{(1-\frac{3M}{2r})}{(1-\frac{2M}{r})(1-\frac{\lambda _{0}}{r})},\quad B(r)=(1-\frac{2M}{r}).
\end{equation}

The calculation for the superpotentials yields the following non-vanishing components for the
\begin{itemize}
\item Landau-Lifshitz prescription 
\begin{equation}\label{U1}
U^{ttx}=\frac{2x}{r^{2}}\frac{[(1-\frac{3\,M}{2\,r})-(1-\frac{\lambda _{0}}{r%
})(1-\frac{2\,M}{r})]}{(1-\frac{2\,M}{r})(1-\frac{\lambda _{0}}{r})}, 
\end{equation}
\begin{equation}\label{U2}
U^{tty}=\frac{2y}{r^{2}}\frac{[(1-\frac{3\,M}{2\,r})-(1-\frac{\lambda _{0}}{r%
})(1-\frac{2\,M}{r})]}{(1-\frac{2\,M}{r})(1-\frac{\lambda _{0}}{r})}, 
\end{equation}
\begin{equation}\label{U3}
U^{ttz}=\frac{2z}{r^{2}}\frac{[(1-\frac{3\,M}{2\,r})-(1-\frac{\lambda _{0}}{r%
})(1-\frac{2\,M}{r})]}{(1-\frac{2\,M}{r})(1-\frac{\lambda _{0}}{r})}. 
\end{equation}
where the $U$'s are defined by Eq.(\ref{LLgauss}).

\item Weinberg prescription
\begin{equation}\label{D1}
D^{xtt}=\frac{2x}{r^{2}}\frac{[(1-\frac{3\,M}{2\,r})-(1-\frac{\lambda _{0}}{r%
})(1-\frac{2\,M}{r})]}{(1-\frac{2\,M}{r})(1-\frac{\lambda _{0}}{r})}, 
\end{equation}
\begin{equation}\label{D2}
D^{ytt}=\frac{2y}{r^{2}}\frac{[(1-\frac{3\,M}{2\,r})-(1-\frac{\lambda _{0}}{r%
})(1-\frac{2\,M}{r})]}{(1-\frac{2\,M}{r})(1-\frac{\lambda _{0}}{r})}, 
\end{equation}
\begin{equation}\label{D3}
D^{ztt}=\frac{2z}{r^{2}}\frac{[(1-\frac{3\,M}{2\,r})-(1-\frac{\lambda _{0}}{r%
})(1-\frac{2\,M}{r})]}{(1-\frac{2\,M}{r})(1-\frac{\lambda _{0}}{r})}. 
\end{equation}
\end{itemize}

Utilizing Eq.(\ref{LLgauss}) along with the metric (\ref{metricSC}) and the quantities
(\ref{U1})-(\ref{U3}) we obtain for the energy distribution  inside a 2-sphere of radius $r$
\begin{equation}\label{LLenergy}
E_{LL}=\frac{r}{2}\frac{[(1-\frac{3\,M}{2\,r})-(1-\frac{\lambda _{0}}{r})(1-%
\frac{2\,M}{r})]}{(1-\frac{2\,M}{r})(1-\frac{\lambda _{0}}{r})}, 
\end{equation}
while the momenta are found to be zero.

In a similar way, using Eq.(\ref{Wgauss}) and (\ref{D1})-(\ref{D3}) we get, for the
Weinberg prescription, an expression for the energy inside a 2-sphere of
radius $r$ that is equal to the one obtained in the Landau-Lifshitz case
\begin{equation}\label{LL-Wenergy}
E_{W}=E_{LL}=\frac{r}{2}\frac{[(1-\frac{3\,M}{2\,r})-(1-\frac{\lambda _{0}}{r})
(1-\frac{2\,M}{r})]}{(1-\frac{2\,M}{r})(1-\frac{\lambda _{0}}{r})},
\end{equation}
while, again, the momenta are found to be zero.

Inserting the specific value $\lambda _{0}=\frac{3\,M}{2}$ in (\ref{LL-Wenergy}), we obtain 
\begin{equation}\label{lambda}
E_{LL}=E_{W}=M(1-\frac{2\,M}{r})^{-1}, 
\end{equation}
which is identical to the expression obtained by Virbhadra for the
Schwarzschild metric \cite{Virb}.

It is seen from (\ref{LL-Wenergy}) that the energy in both the Landau-Lifshitz and the Weinberg prescriptions 
depends on the mass $M$ of the black hole, the parameter $\lambda _{0}$  and the radial coordinate $r$.

\section{Discussion}

In our paper we have studied the energy-momentum for a new black hole solution in the brane world described by the line element 
\[
ds^{2}=(1-\frac{2M}{r})dt^{2}-\frac{(1-\frac{3M}{2r})}{(1-\frac{2M}{r})(1-%
\frac{\lambda _{0}}{r})}dr^{2}-r^{2}(d\theta ^{2}+\sin ^{2}\theta d\varphi
^{2}).
\] 
In particular, we have used the
definitions of Landau-Lifshitz and Weinberg in order to calculate the
expressions for the energy and the momentum. It is found that, in the gravitational background considered, all the momenta vanish while the energy depends on the mass $M$ , the radial coordinate $r$ and the positive parameter 
$\lambda_{0}$, for both the Landau-Lifshitz and Weinberg prescriptions. In fact, the energy obtained is found to be the same in both prescriptions and it is given by 
\[
E_{W}=E_{LL}=\frac{r}{2}\frac{[(1-\frac{3\,M}{2\,r})-(1-\frac{\lambda _{0}}{r})
(1-\frac{2\,M}{r})]}{(1-\frac{2\,M}{r})(1-\frac{\lambda _{0}}{r})}.
\]
Specifically, for the value $\lambda _{0}=\frac{3\,M}{2}$ the expression for the energy reads $E_{LL}=E_{W}=M(1-\frac{2\,M}{r})^{-1}$, being the result obtained by Virbhadra for the Schwarzschild metric \cite{Virb}. As it is evident,  for $r\rightarrow \infty $ the last expression gives $E_{LL}=E_{W}=M$ which represents the ADM mass, while for $r\rightarrow 0$, $E_{LL}=E_{W}=0$.

In the following table we summarize the limiting behavior of the energy in both prescriptions for two particular cases:

\[
\begin{tabular}{ccc}
Limit & Energy$_{LL}$ & Energy$_{W}$ \\ 
\hline
\\
$r\rightarrow 0$ & $0$ & $0$ \\ 
&&\\
$r\rightarrow 2M$ & $\infty$ & $\infty$ \\ 
&&\\
$r\rightarrow \infty $ & $\frac{1}{2}(\frac{M}{2}+\lambda _{0})$ & $\frac{1}{%
2}(\frac{M}{2}+\lambda _{0})$\\
&&
\end{tabular}%
\]

At $r=2M$ lies the event horizon of the black hole as in the Schwarzschild case. The result $E_{LL}=E_{W}=\frac{1}{2}(\frac{M}{2}+\lambda_{0})$ for $r\rightarrow\infty$ was also obtained by Gamal G.L. Nashed \cite{Gamal_2} in the context of TEGR. If we consider this limiting case for $\lambda _{0}=\frac{3\,M}{2}$ we get again $E=M$ which is consistent with the energy obtained for the Schwarzschild solution, whereby $M$ is the ADM mass. The agreement between our result and the result obtained in the context of TEGR  is worth to be pointed out 
if one considers that general relativity is a geometrical theory (the metric tensor being its fundamental field), while TEGR is actually a gauge theory (with a gauge potential as its fundamental field) \cite{Aldro}. Thus, we have come up with the same  result not only by using a different theory but also by applying a different methodology.

Even if our present work does not settle the problem of the energy-momentum localization of the gravitational field by using energy-momentum complexes, the results obtained can be considered as a contribution to the ongoing debate concerning this issue.

\section*{Acknowledgements}

The authors would like to thank  Dr. G. O. Papadopoulos for his valuable suggestions.

\end{document}